\documentclass[12]{article}
\usepackage{amsmath,amssymb,amsthm}
\usepackage{mathtools}
\usepackage{mathrsfs}
\usepackage{bm}
\usepackage{slashed} 
\usepackage{multirow}
\usepackage{tikz}
\usepackage[utf8]{inputenc}
\usepackage{fullpage}
\usepackage{arydshln}
\usepackage{setspace}
\usepackage[margin=1in]{geometry}
\usepackage{dsfont}
\usepackage{graphicx ,lipsum}
\graphicspath{ {graphs/} }
\usepackage{subcaption}
\usepackage{float}
\usepackage{xcolor}
\usepackage{cite}
\usepackage{subcaption}
\usepackage{mwe}

\usepackage{hyperref}
\usepackage[symbol]{footmisc}

\hypersetup{colorlinks=true,linkcolor=blue,anchorcolor=blue,citecolor=green,filecolor=blue
urlcolor=blue,bookmarksnumbered=true,pdfview=FitB}

\usepackage{epstopdf}

\begin{document}
\begin{center}
{\Large \bf \strut
Coulomb Excitation of Deuteron in Peripheral Collisions with a Heavy Ion
\strut}\\
\vspace{10mm}
{\large \bf 
Weijie Du$^{a}$, Peng Yin$^{b,a}$\footnote[1]{Corresponding author: yinpeng@impcas.ac.cn}, Yang Li$^{c}$, Guangyao Chen$^{a,d}$, 
Wei Zuo$^{b,e}$, Xingbo Zhao$^b$ and James P. Vary$^{a}$} 

\end{center}

\noindent{\small $^a$ \it{Physics Department, Iowa State University, Ames, Iowa, U.S.A., 50010}} \\
{\small $^b$\it Institute of Modern Physics, Chinese Academy of Sciences, Lanzhou, China, 730000} \\
{\small $^c$ \it{Department of Physics, College of William and Mary, Williamsburg, Virginia, U.S.A., 23168}}\\
{\small $^d$ \it{Department of Physical Sciences, Perimeter College, Georgia State University, Alpharetta, Georgia, U.S.A., 30022}}
{\small $^e$ \it{University of Chinese Academy of Sciences, Beijing, China, 100049} }

\section*{Abstract}
We develop an {\it ab initio}, non-perturbative, time-dependent Basis Function (tBF) method to solve the nuclear structure and scattering problems in a unified manner. We apply this method to a test problem: the Coulomb excitation of a trapped deuteron by an impinging heavy ion. The states of the deuteron system are obtained by the {\it ab initio} nuclear structure calculation implementing a realistic inter-nucleon interaction with a weak external trap to localize the center of mass and to discretize the continuum. The evolution of the internal state of the deuteron system is directly solved using the equation of motion for the scattering. We analyze the excitation mechanism of the deuteron system by investigating its internal transition probabilities and observables as functions of the exposure time and the incident speed. In this investigation, the dynamics of the Coulomb excitation are revealed by the time evolution of the system's internal charge distribution.

\section{Introduction}
A unified treatment of nuclear structure and reactions is a central, but challenging, issue of {\it ab initio} nuclear theory. Specifically, the challenge is to incorporate the discrete bound states with the scattering states in the continuum \cite{Feshbach:1958nx,Feshbach:1962ut}. For few-body systems with mass number $A \leq 4$, highly precise  methods such as Faddeev \cite{Witala:2000am}, Faddeev-Yakubovsky \cite{Lazauskas:2004hq,Lazauskas:2009gv}, Alt-Grassberger and Sandhas \cite{Deltuva:2006sz,Deltuva:2007xv}, and hyperspherical harmonics \cite{Kievsky:2008es,Marcucci:2009xf} have been developed using internal coordinates. As for light and medium nuclei with $A>4$, there are also a wealth of cutting edge approaches. A survey of the methods includes the no-core shell model with resonating group method \cite{Quaglioni:2008sm,Quaglioni:2009mn,Navratil:2009ut,Navratil:2011zs}, the no-core shell model with continuum method \cite{Baroni:2012su,Baroni:2013fe,Navratil:2016ycn}, the coupled cluster method with the Gamow basis \cite{Hagen:2010zz,Hagen:2012sh,Hagen:2012rq}, the no-core shell model with the Gamow basis \cite{Papadimitriou:2011jx,Papadimitriou:2013ix,Barrett:2015wza}, the HORSE (J-matrix) method \cite{Bang:2000,Shirokov:2016thl,Shirokov:2016ywq}, the configuration interaction with resonating group method \cite{Kravvaris:2017nyj}, the Green's function Monte Carlo method \cite{Lynn:2015jua,Nollett:2006su}, and the nuclear lattice effective field theory \cite{Rupak:2013aue,Elhatisari:2015iga}. However, these successful methods may be challenged to retain the full, non-perturbative quantum coherence of the scattering over all potentially relevant intermediate and final states which could be important for complex scattering processes involving exotic nuclei. For short-lived rare isotopes, where the low-lying states are either weakly bound or unbound, one will be challenged to include the relevant degrees of freedom for a complete description of the inelastic processes. In particular, a large number of intermediate states may be needed to provide accurate descriptions of the dynamical multi-step processes contributing to the final states. 

In order to address these complex processes and retain predictive power, we propose an {\it ab initio}, time-dependent non-perturbative approach, which we call the time-dependent Basis Function (tBF) approach. The idea, which is based on a successful time-dependent approach in quantum field theory \cite{Vary:2009gt,Zhao:2013cma,Zhao:2013vga,Chen:2017uuq,Du:2017ckx}, is to solve the equation of motion (EOM) for the scattering of the system in the representation constructed from the energy eigenbases of the system before scattering. The state vector for the system hence reduces to a set of amplitudes with respect to the chosen eigenbases, in which the full coherence is retained, and the EOM becomes a set of first order differential equations in time. 

We demonstrate the tBF approach with a very simple problem, the internal excitation of a trapped deuteron in the time-varying external Coulomb field of a heavy ion, or deuteron Coulomb excitation \cite{Alder:1956im,Winther:1979zz}. Note in this initial application, the motion of the center of mass (COM) of the deuteron is constrained to the trap and the excitation in the COM degree of freedom is neglected. Future work will remove the trap and evolve the motion of the COM. Within the tBF formalism, the evolution of the deuteron system is examined through its consequent transition probabilities and through expectation values of different observables during the scattering. The dynamics of the scattering process will also be revealed by the time evolution of the deuteron system's internal charge density distribution. 

This paper is organized as follows. We first introduce the theory of the tBF approach in Sec. 2. Then, we discuss the details of our model problem in Sec. 3 and present the simulation conditions of the problem in Sec. 4. Later, we provide illustrative numerical results in Sec. 5. Finally, we present conclusions and outlook in Sec. 6. The appendix contains useful mathematical details of the spherical harmonic oscillator basis.

\section{Theory of the tBF approach}
We begin with an introduction of the framework for time-dependent scattering within a basis space determined from an {\it ab initio} structure calculation. In particular, we outline the problem where the external field, which induces the transitions, is treated as a classical, possibly strong, time-dependent electromagnetic (EM) source. The generalization to more complex sources will be considered in subsequent works. To be concrete and simple, we outline the approach for the specific case of a trapped deuteron as the system undergoing excitation, which, however, can be straightforwardly generalized.

\subsection{Hamiltonian}
Our full Hamiltonian for the target scattered by the time-varying EM field produced by the impinging heavy ion (HI) is
\begin{eqnarray}
H_{\text{full}}(t) &=& H_0 + V_{\text{int}}(t) \label{eq:FullH} \ ,
\end{eqnarray}
where the Hamiltonian for the intrinsic motion of the target is 
\begin{eqnarray}
H_0 &=& T_{\text{rel}} + V_{\text{NN}} + U_{\text{trap}} \label{eq:H0} \ ,
\end{eqnarray}
with $T_{\text{rel}}$ the relative kinetic energy and $V_{\text{NN}}$ the nucleon-nucleon ($NN$) interaction. $U_{\text{trap}}$ denotes an external harmonic oscillator (HO) trap introduced to localize the COM of the target and to discretize the continuum of the target's scattering states. We neglect the excitation of the COM motion. Removing the regularization provided by the trap will be the subject of future investigations. 

For physical motivation to retain a weak trap, one may cite the utility of a quasi-deuteron approach to reactions as an example. In that case, the presence of our trap simulates a nuclear environment in which the deuteron degree of freedom is selected to respond to an external probe \cite{JEisenbergV587}. 

The time-dependent interaction between the target and the external EM field is $V_{\text{int}}(t)$, which is formulated by the coupling between the four current $J^{\mu}=(\rho,\ \vec{j})$ of the target and the four potential $A^{\mu}=(\varphi,\ \vec{A})$ of the external EM field 
\begin{eqnarray}
V_{\text{int}}(t) &=& \int A_{\mu}J^{\mu} \ d \vec{r} \ = \ \int \rho (\vec{r},t) \varphi(\vec{r},t) \ d \vec{r} \ -\ \int \vec{j}(\vec{r},t) \cdot \vec{A}(\vec{r},t) \ d \vec{r} \ \label{eq:CouplingEnergy} \ .
\end{eqnarray}
Note we adopt the natural units and set $\hbar = c = 1$ throughout this paper. 

\subsection{EOM for the scattering}
The EOM for the target during the scattering, in the interaction picture, is 
\begin{eqnarray}
i \frac{\partial}{\partial t}|\psi; t \rangle _I &=&  e^{i {H_{0}t}}\ V_{\text{int}}(t)\ e^{-i {H_{0}}t}\ |\psi; t \rangle _I \ \equiv \  V_I(t)\ |\psi; t \rangle _I \ , \label{eq:EOMequation}
\end{eqnarray}
where $V_I(t)$ denotes the interaction part in the full Hamiltonian. The subscript ``I'' specifies the interaction picture. The state vector of the target at time $t \ge t_0$ ($t_0$ is the time when the target is defined in its initial state and begins to experience the time-dependent interaction) can be solved as
\begin{eqnarray}
|\psi; \ t \rangle _I &=& U_I(t;t_0) |\psi;\ t_0 \rangle _I \ \label{eq:EOMsoln} \ ,
\end{eqnarray} 
where $U_I(t;t_0)$ is the unitary operator for the time-evolution
\begin{eqnarray}
U_I(t;t_0) &=& \hat{T} \Bigg\{  \exp \Bigg[-i\int_{t_0}^t \ V_I(t')\ dt' \Bigg]\Bigg\} \label{eq:timeEvlveOperator} \ ,
\end{eqnarray}
with $\hat{T}$ the time-ordering operator towards the future. 

The time-evolution operator $U_I(t;t_0)$ can be evaluated numerically by first dividing the interval $[t_0,t]$ into segments with step length $\delta t = (t-t_0)/n$ ($n$ being sufficiently large to attain numerically stable results) and then replacing the integration in the exponent with additive increments. Keeping only terms up to the order of $\delta t$ in the following Taylor expansion, we get
\begin{align}
U_I(t;t_0) \xrightarrow{\sum \delta t} \Bigg[ 1-i\ V_I(t)\delta t \Bigg]\ \Bigg[ 1-i\ V_I(t_{n-1})\delta t \Bigg] \ \cdots \ \Bigg[ 1-i\ V_I(t_1)\delta t \Bigg] \label{eq:Evolv} \ .
\end{align}
The direct evaluation according to Eq. (\ref{eq:Evolv}) is called the Euler scheme. It is numerically unstable since this scheme is not symmetric in time; the norm of the state vector of the target may not be conserved \cite{AAskar78} during the evolution. We therefore adopt the MSD2 scheme \cite{TIitaka94} in our tBF method. Via the MSD2 scheme, the state vector for the target at the time $t_1=t_0+ \delta t$ is still evolved via the Euler scheme. However, for $t'=t_2, t_3, \cdots, t_{n-1}$, the state vector under time evolution is
\begin{eqnarray}
|\psi; t'+\delta t \rangle _I &\approx & |\psi; t'-\delta t\rangle _I -{2i} \ V_I(t') \ \delta t \ |\psi; t'\rangle _I \label{MSD2} \ .
\end{eqnarray}
For the current model problem, we also calculate the state vector of the target via first-order perturbation theory for comparison
\begin{eqnarray}
|\psi; \ t\rangle _I  &\rightarrow&  \Bigg[1 -i\ \Big(V_I(t)+ \cdots V_I(t_2)+V_I(t_1) \Big) \delta t  \Bigg] |\psi;\ t_0\rangle _I \ ,
\end{eqnarray}
where only the terms up to the order of $\delta t$ are retained. 

\subsection{Basis representation}
We solve the energy eigenbases of the target from its intrinsic Hamiltonian (Eq. \eqref{eq:H0}). The eigenequation is
\begin{eqnarray}
H_{0} |\beta_j \rangle &=& E_{j} \ |\beta_j \rangle \label{eq:EFunc} ,
\end{eqnarray}
where $E_j$ is the eigenvalue corresponding to the eigenvector $| \beta_j \rangle$ and the subscript $j$ is an index running over the individual states. In the basis representation defined by the set of bases $\{|\beta_j \rangle \}$, the state vector of the target becomes a vector of time-dependent amplitudes, while the operators become matrices and the EOM (Eq. \eqref{eq:EOMequation}) becomes sequential matrix-vector multiplications.

\subsection{Transition amplitude}
In the basis representation, the state vector of the target at any moment $t$ during the scattering is 
\begin{eqnarray}
| \psi ; t \rangle _I &=& \sum_j A^I_j(t) | \beta _j \rangle \ \label{eq:stateVectorSoln1} ,
\end{eqnarray}
where the $A^I_j(t)$ is the amplitude corresponding to the basis $| \beta _j \rangle$. Given the initial state vector of the target at the beginning of the scattering ($t=t_0$) to be $| \psi ; t_0 \rangle =  | \beta _i \rangle $, $A^I_j(t)$ describes the transition amplitude from $| \beta _i \rangle $ to $| \beta _j \rangle $ and can be computed as
\begin{eqnarray}
A^I_j(t) &=& \langle \beta _j | U_I(t;t_0) | \beta _i \rangle \ \label{eq:stateVectorSoln2} ,
\end{eqnarray} 
with $A_j^I(t_0)= \delta _{ij}$.
The corresponding transition amplitude in the Schr\"odinger picture is
\begin{eqnarray}
A_j(t) &=& \exp \Bigg[- i E_j t + i E_i t_0  \Bigg] A^I_j(t) \ ,
\end{eqnarray}
and the full state vector of the target at time $t$ is
\begin{eqnarray}
| \psi ; t \rangle &=& \sum_j A_j(t) | \beta _j \rangle \ .
\end{eqnarray}

\subsection{Observables and the density distribution}
Based on $| \psi ;t \rangle $, we can calculate the expectation values of observables during the scattering as 
\begin{eqnarray}
\langle O(t) \rangle &=&  \langle \psi; \ t |\ \hat{O} \ | \psi; \ t \rangle  \ = \ \sum_{j,k} A_j^{\ast}(t) A_k(t) \ \langle \beta_j | \hat{O} | \beta_k \rangle  \ , \label{eq:MatrixRepresentationOfOperators}
\end{eqnarray}
where $\hat{O}$ denotes the operator for the selected observable. 

As an example, we can study the dynamics of the target via the evolution of its effective charge density distribution, which is formulated as
\begin{eqnarray}
\rho (\vec{r};t) &=& \langle \psi ; t | \vec{r} \rangle \langle \vec{r} | \psi ; t \rangle \ = \ \sum_{jk} A_k^{\ast}(t) A_j(t) \langle \beta _k | \vec{r} \rangle \langle \vec{r} | \beta_j \rangle \ , \label{eq:densityR} 
\end{eqnarray}
where $\langle \vec{r} | \beta_j \rangle$ denotes the wave function of the $j^{th}$ basis in coordinate space. The charge density distribution of the target in its relative coordinates will be simply referred to as the internal charge distribution in the following text.

\section{Setup of the model problem}

\begin{figure}[H]
\centering
\includegraphics[width=14cm]{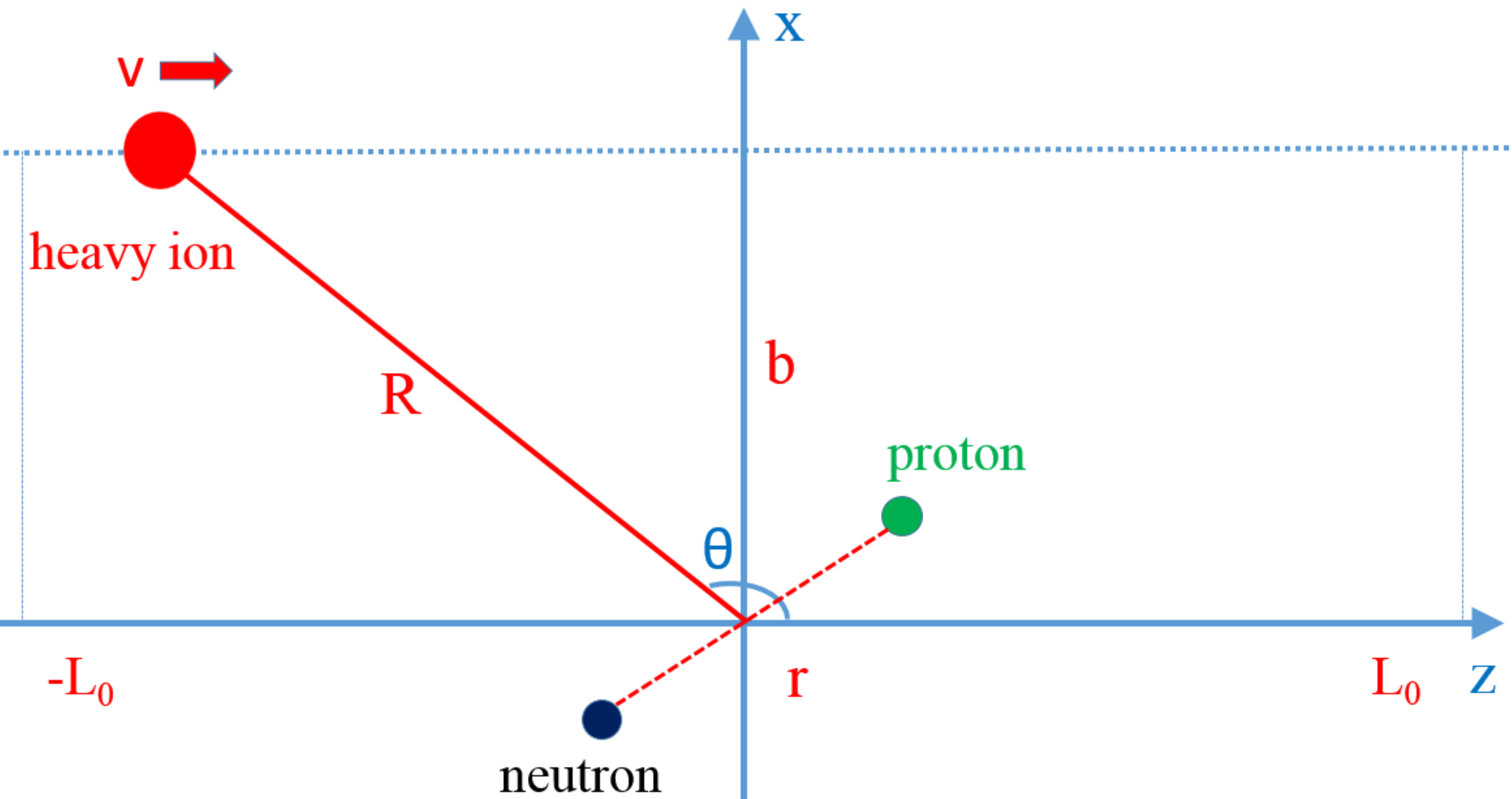}
\caption{Set up of the peripheral scattering (adopted from Ref. \cite{Du:2017ckx}). See the text for the details.}
\label{fig:MoonAnd7Pennies}
\end{figure}

As shown in Fig. \ref{fig:MoonAnd7Pennies}, we set the scattering plane to be the $xz$-plane. The target is a deuteron. For simplicity, we assume its COM is fixed at the origin, i.e., the recoil of the target during the scattering is neglected. The relative coordinates of the target are defined as $\vec{r}=\vec{r}_p - \vec{r}_n$, where $\vec{r}_p$ and $\vec{r}_n$ are the single-particle coordinates for the proton and the neutron, respectively. The masses of the neutron and the proton are taken to be their average mass 938.92 MeV. The mesonic degree of freedom is not considered and the unit charge of the target is carried by the proton. 

The projectile is a HI. It carries charge $Ze$ and is assumed to move, for simplicity here, with a constant velocity $\vec{v}$ parallel to the $\hat{z}$-axis. The impact parameter $b$ is set to be sufficiently large such that the nuclear interaction is negligible compared to the EM interaction during the scattering. $\vec{R}$ denotes the position of the HI with respect to the origin. 

\subsection{Background field}
As an initial application of the tBF method, we assume that the HI impinges with a low speed (non-relativistic) and the magnetic interaction between the target and the induced vector field $\vec{A}(\vec{r},t)$ is ignored. That is, we evaluate only the interaction between the target and the time-varying Coulomb field. We then perform the multipole expansion of the Coulomb field \cite{ABohr} and, for this initial application, we retain only the contribution of the $E1$ multipole component. The investigation on the contributions of other components (e.g., $E0$, $E2$) as well as the magnetic transitions (e.g., $M1$) will be addressed in the future. 

In the basis representation, the operator for the $E1$ multipole component \cite{Alder:1956im,JEisenberg88} of the time-varying Coulomb interaction $V_I(t)$ becomes a matrix with elements formulated as 
\begin{eqnarray}
\langle \beta _j | V_I(t) | \beta _k \rangle &=& \frac{4 \pi}{3} Z e^2 e^{i({E_{j} - E_{k}})t} \sum_{\mu} \ \frac{Y^{\ast}_{1 \mu}(\Omega_{{R}})}{|R(t) |^{2}} \int d \vec{r} \ \langle \beta _j | \vec{r} \rangle \ \frac{r}{2}Y_{1 \mu}(\Omega_{{r}}) \ \langle \vec{r} | \beta _k \rangle \label{eq:IntMatElem} \ ,
\end{eqnarray}
where $Y_{\lambda \mu}(\Omega)$ denotes the spherical harmonics (the Condon-Shortley convention \cite{JSuhonen} is adopted in this work). $\lambda =1$ denotes the dipole contribution out of the multipole components of the Coulomb field. $\Omega _R$ denotes the direction of the HI, which is specified by the polar angle and the azimuth angle of $\vec{R}$. Similarly, $\Omega_{r}$ is specified by the polar and azimuth angles of $\vec{r}$. The matrix representation for the time-evolution operator $U_I(t;t_0)$ can thus be solved according to Eq. \eqref{eq:IntMatElem}.

\subsection{Structure calculation of the target}
In our tBF method, we solve for target properties by an {\it ab initio} nuclear structure calculation. In this work, the three dimensional (spherical) harmonic oscillator (3DHO) representation in relative coordinates is implemented to calculate the eigenenergies and the corresponding eigenbases. For the internal motion of the deuteron system, each 3DHO basis $|nlSJM \rangle $ is specified by the radial quantum number $n$, the quantum number $l$ for the orbital angular momentum, the quantum number $S$ for the spin, the quantum number $J$ for the total angular momentum (we adopt the scheme where $l$ is coupled to $S$ to form $J$) and the magnetic quantum number $M$ for the $\hat{z}$-projection of the total angular momentum. The truncation parameter for the model space is defined by $2n+l \le N_{\text{max}}$. Hence the model's 3DHO basis set $\{ |nlSJM \rangle \} $ is specified by good quantum numbers $S$, $J$, $M$ and parity (determined by $(-1)^l$) of the $np$ system. We thus define our retained eigenbasis in Eq. \eqref{eq:EFunc} as 
\begin{eqnarray}
| \beta _j  \rangle &=& \sum_{{n_j l_j}} a _{n_j l_j} | n_j l_j S_j J_j M_j \rangle \ \label{eq:EigenBasis3DHORep} ,
\end{eqnarray}
where $\beta$ stands for $l$, $S$, $J$ and $M$ for each channel. $\{a _{n_j l_j} \}$ denotes the set of the expansion coefficients, which are obtained by the diagonalization of the matrix $H_0$ in the 3DHO representation. The kernel in Eq. \eqref{eq:MatrixRepresentationOfOperators} thus becomes
\begin{eqnarray}
\langle \beta_j | \hat{O} | \beta_k \rangle &=& \sum_{n_jl_j} \sum_{n_kl_k}  a^{\ast} _{n_j l_j} a_{n_kl_k} \langle n_jl_jS_jJ_jM_j | \hat{O} | n_kl_kS_kJ_k M_k \rangle \ \label{eq:3DHOOperator} .
\end{eqnarray}
Details of our conventions for the 3DHO basis representation, the EM operators and the observables employed here in the 3DHO basis are presented in the appendix.

\section{Simulation conditions}
In this work, we will adopt a concrete but simple test application to demonstrate the feasibility of the tBF method and to gain an initial appreciation of the coherent features available in time-dependent evolution at the amplitude level. The projectile is taken as a fully stripped uranium, U$^{92+}$. The incident speeds are set to be 0.1, 0.2 and 0.4. We fix the duration of exposure time to be from $-5$ MeV$^{-1}$ to $5$ MeV$^{-1}$, which is approximately 6.582$\times 10^{-21}$ sec. The impact parameter is chosen as $b=5$ fm. That is, as an example, the projectile with the incident speed $v=0.1$ travels from 100 fm before the distance of the closest approach between the projectile and the origin to 100 fm after the closest approach. 

\begin{figure}[H]
\centering
\includegraphics[width=8cm]{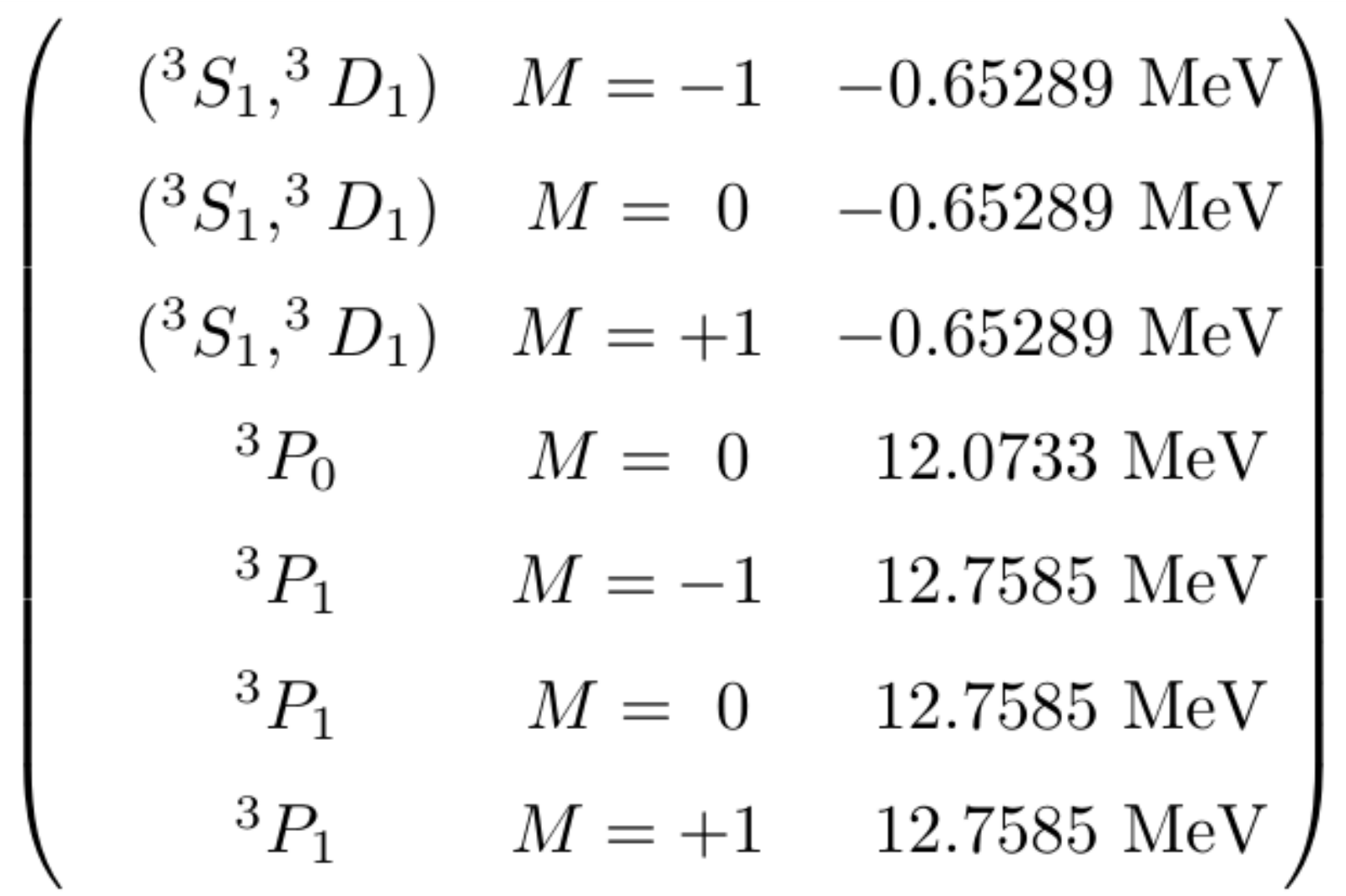}
\caption{The eigenbasis vector of the target deuteron confined in an external HO trap of strength $5$ MeV. This vector makes explicit the basis representation for our model and lists the channel quantum numbers, the angular momentum projection, and the eigenenergies. For the {\it ab initio} structure calculation, the 3DHO bases are adopted, for which the basis strength is set to be $\omega = 5$ MeV and the truncation parameter $N_{\text{max}}$ to be 60. The lowest-lying 7 states are chosen to construct the basis representation for the target. Note there are the expected degeneracies with respect to the target system's magnetic projection $M$.}
\label{fig:SevenBases}
\end{figure}

One of the main features of the tBF approach is the ability to incorporate microscopic  nuclear structure via the {\it ab initio} method with an adopted realistic nuclear interaction. For the current work, we adopt the JISP16 \cite{AShi04, AShi05, AShi07} realistic $NN$-interaction to construct the target Hamiltonian (Eq. \eqref{eq:H0}). In the 3DHO representation, the eigenenergies and the corresponding eigenstates of the $np$ target are solved according to Eq. \eqref{eq:EFunc} with both the trap and basis strengths taken to be 5 MeV and $N_{\text{max}}=60$. For simplicity, we take only three interaction channels for the target, which are $(\ ^3S_1,\ ^3D_1)$, $\ ^3P_0$ and $\ ^3P_1$. The lowest states of each channel, as shown in Fig. \ref{fig:SevenBases}, are taken into account. In applying the tBF method to this simple demonstration problem, we construct the basis representation for the total time-dependence of the target in terms of these states. The initial state of the target is taken to be $(\ ^3S_1,\ ^3D_1)$, $M=-1$, which is polarized against the $\hat{z}$-axis. More interaction channels, different $NN$-interactions and different targets  will be studied in the future work. 

The interaction between the target and the time-varying external Coulomb field is then expressed as matrix elements in the basis representation. According to the equation of motion (Eq. \eqref{eq:EOMequation}), the time-dependent state vector of the target can be solved in the form of Eq. \eqref{eq:stateVectorSoln1} and Eq. \eqref{eq:stateVectorSoln2}. 

In this work, we investigate selected observables of the target, the transition probability, the r.m.s. charge radius, the r.m.s. intrinsic momentum, the r.m.s. angular momentum, the intrinsic energy and the $\hat{z}$-projection of the total angular momentum, as functions of the exposure time and the incident speed (or, equivalently, bombarding energy). To help formulate our intuition, we also present some details of the evolution of the internal charge distribution (Eq. \eqref{eq:densityR}) during the scattering. 

\section{Results and discussions}

\subsection{Transition probabilities}

\begin{figure}[H]
\centering
\includegraphics[width=16cm]{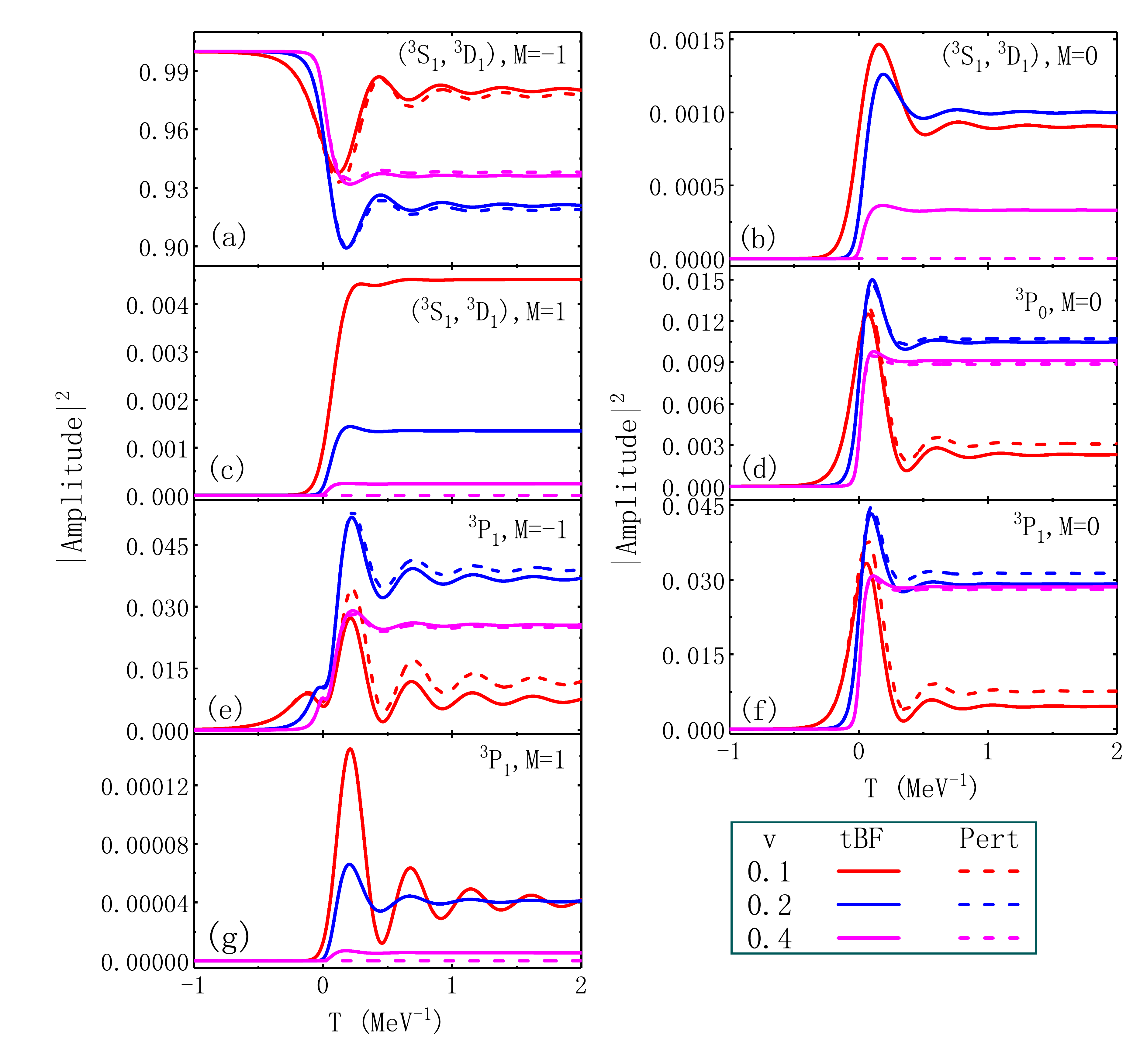}
\caption{The Coulomb excitation (only the $E1$ multipole component is included) of the  target illustrated as a function of the time and the incident speed of the HI in the middle of the scattering. The target is characterized by 7 basis states. It is initially prepared to be in the state $(^3S_1,\ ^3D_1),\ M=-1$. The HI projectile is taken to be a fully stripped uranium, U$^{92+}$, and the incident speed is taken as 0.1, 0.2 and 0.4. The transition probabilities of each basis state of the target are calculated via the non-perturbative tBF method and compared with results from first-order perturbation theory (curves labeled by ``Pert" in the legend). }
\label{fig:TransitionProb}
\end{figure}

With the total exposure time fixed and only the incident speed altered, we present in Fig. \ref{fig:TransitionProb} the transition probabilities of the basis states of the $np$ target as functions of the time and the incident speed of the HI at intermediate times (from $-1$ MeV$^{-1}$ to $2$ MeV$^{-1}$), which covers the time period where the significant transitions occur. Note we ignore the corrections from the relativistic effects and the magnetic transitions caused by the induced vector field. For the numerical calculation, we apply the same method introduced in our previous paper \cite{Du:2017ckx}, where we checked that the tBF method agrees with first-order perturbation theory when the external Coulomb field is sufficiently weak and the first-order effects dominate. In addition, we conduct two more validity checks. First, the normalization of the time-dependent wave function is verified during the evolution of the $np$ target. Second, the time-reversal symmetry of the algorithm for the evolution is verified by explicitly running the solution backwards to the initial state. 

\subsubsection{General features of the excitation}
During the scattering, when the HI projectile is sufficiently close to the mass center of the target, abrupt transitions occur and the probabilities exhibit short-time fluctuations. Such quantum fluctuations are expected in the quantal treatment of scattering and we verified that these quantum fluctuations are consistent with the uncertainty relation. We clearly observe such quantum fluctuations in, for example, the evolution of the initial state $(^3S_1,\ ^3D_1),\ M=-1$ with an incident speed $v=0.4$ in Fig. \ref{fig:TransitionProb}. Here, the elastic scattering probability dips sharply and relaxes to its asymptotic value. For this case, the full width half maximum (FWHM) (for the first dip during the evolution) is $\Delta t> 0.1 $ MeV$^{-1}$, while the transition energy is $\Delta E>12.7$ MeV, yielding a product greater than unity which is consistent with the uncertainty principle. 

Eventually, short-time fluctuations attenuate and approach asymptotic values  as the Coulomb field fades away. The excited target then evolves into a final superposition of the available eigenstates of the target Hamiltonian. From the produced scattering amplitude at later times, the amplitude for breakup into a particular, kinematically allowed, final state is found by projecting onto that final state. In reality, the excited target can also decay through other kinematically accessible
channels, such as through spontaneous EM radiation, which is not included in the present model.

\subsubsection{Allowed and forbidden transitions}
In Fig. \ref{fig:TransitionProb}, the difference in the transition probabilities given by the non-perturbative tBF method and the corresponding first-order perturbation theory shows the importance of the higher-order effects during the scattering process. Specifically, since only the $E1$ multipole component of the time-varying Coulomb field is included, we expect the dominant transitions in Fig. \ref{fig:TransitionProb} to reflect the $E1$-selection rule for the calculations based on first-order perturbation theory. We refer to transitions from the initial state that are permitted by first-order perturbation theory as ``allowed" and all other transitions as ``forbidden" for the purposes of this discussion. However, for the current setup ($Z=92,\ b=5$ fm), the Coulomb interaction is strong when the projectile is close to the target; higher-order effects, which are included by the non-perturbative tBF approach, produce some major consequences when compared with first-order perturbation theory. For example, first-order perturbation theory predicts $(^3S_1,\ ^3D_1),\ M=1$ to be a ``dark" state (an $E1$-forbidden transition), while its population is clearly revealed by the non-perturbative tBF method during the scattering process via a succession of $E1$ transitions through the accessible intermediate states. The tBF population of two additional dark states is shown in Fig. \ref{fig:TransitionProb}, which is evident by the contrasting null results from first-order perturbation theory.  For the allowed transitions in Fig. \ref{fig:TransitionProb}, there are visible differences in the magnitudes between the tBF and the perturbation theory results with first-order perturbation theory tending to overestimate the transition probability for the simulations with the incident speed 0.1 and 0.2.

\begin{figure}[H]
\centering
\includegraphics[width=12cm]{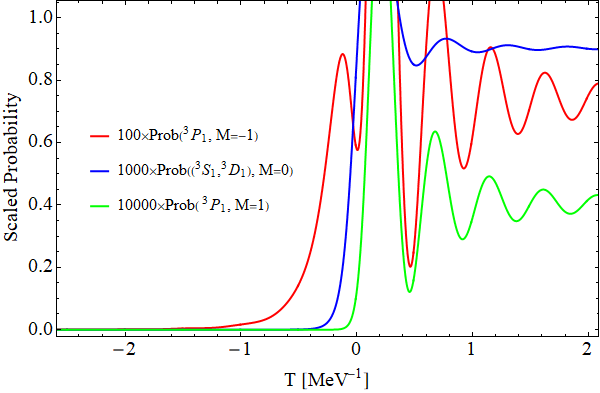}
\caption{Illustration of the state populations changing with time during the scattering.  Note the transition probabilities are all scaled. The $E1$-allowed state $^3P_1,\ M=-1$ is populated initially. Then one observes the transport of population from the state $^3P_1,\ M=-1$ to the $1^{\text{st}}$-order forbidden state $(^3S_1,\ ^3D_1),\ M=0$. Later, the transition network populates the $2^{\text{nd}}$-order forbidden state $^3P_1,\ M=1$ with that population fed from the state $(^3S_1,\ ^3D_1),\ M=0$. The forbidden states also receive population from other states, in which cases relative phases  can lead to interference. }
\label{fig:plotProbTimeBehavior}
\end{figure}

The time sequence of the transition probabilities is illustrated in Fig. \ref{fig:plotProbTimeBehavior}. The states that obey the $E1$-selection rule from the ground state are populated earlier with more population (e.g., $ ^3P_1,\ M=-1$), compared to transitions forbidden at leading order. Shortly thereafter, secondary transitions begin to populate the $1^{\text{st}}$-order forbidden states (e.g., $(^3S_1,\ ^3D_1),\ M=0$) (these secondary routes are referred to as populating the ``1st-order forbidden" states). However, these effects do not significantly populate the forbidden states until the $E1$-allowed states accumulate appreciable population. It is important to note that the de-excitation of states is also included among the transitions. After the $1^{\text{st}}$-order forbidden states are sufficiently populated, the transition network starts to build up the population for the $2^{\text{nd}}$-order forbidden states, e.g., $^3P_1,\ M=1$. In general, the forbidden states populated by the higher-order transitions build up relatively smaller populations. 

\subsubsection{Dependencies of the transitions on the incident speed}
\begin{figure}[H]
\centering
\includegraphics[width=16cm]{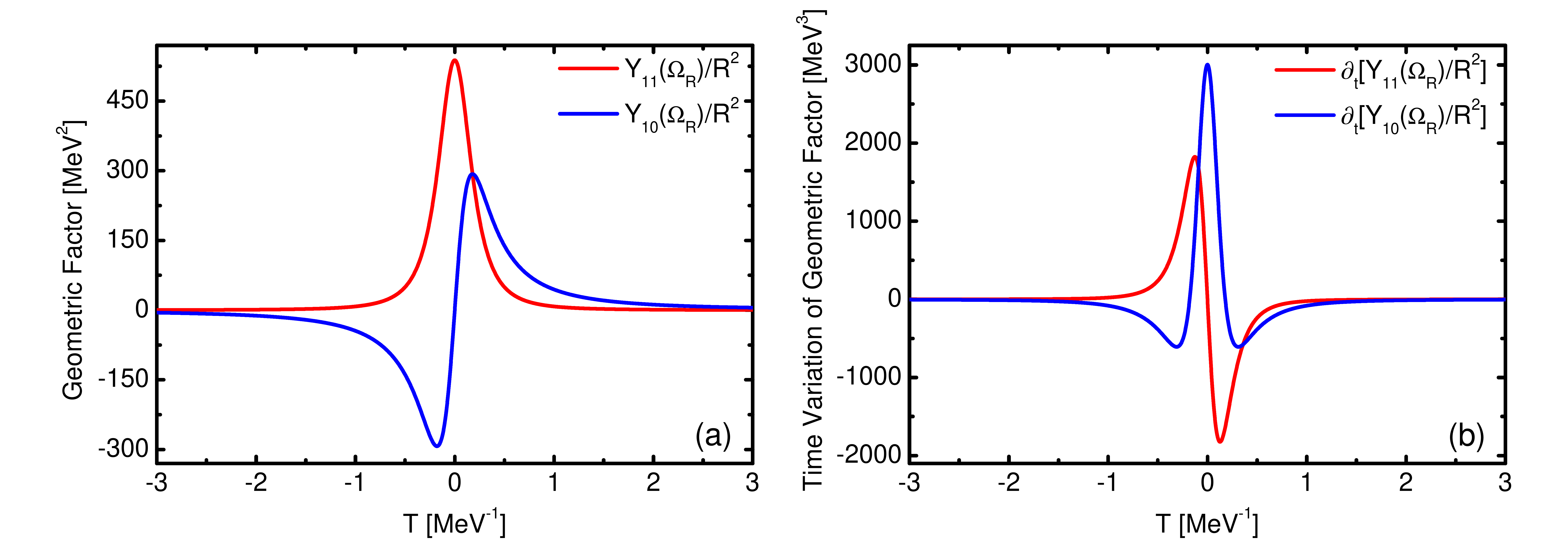}
\caption{Illustration of the geometric factor (panel (a)) $\frac{Y^{\ast}_{1 \mu}(\Omega_{{R}})}{|R(t) |^{2}}$ ($\mu = \pm 1,\ 0$) and its time-variation (panel (b)) during the scattering. The incident speed is taken as $v=0.1$. 
Since the scattering plane is the $xz$-plane, the azimuth angle vanishes and $Y^{\ast}_{1 \mu}(\Omega_{{R}})$ is real. The results related to $Y_{1-1}(\Omega_{{R}})$ are omitted owing to the fact that $Y^{\ast}_{11}(\Omega_{{R}})=-Y_{1-1}(\Omega_{{R}})$.} 
\label{fig:GeoFactor}
\end{figure}

With increasing incident speed, we find that the transitions begin later and that the oscillations of the transition probabilities attenuate more rapidly (transitions experience damping of their oscillatory patterns and approach to asymptotic values). These behaviors can be understood  based on the strength and time-variation of the Coulomb interaction sensed by the target. According to Eq. \eqref{eq:IntMatElem}, the time-variation of the interaction matrix element is, in part, scaled by the geometric factor $\frac{Y^{\ast}_{1 \mu}(\Omega_{{R}})}{|R(t) |^{2}}$. Since we set the scattering plane to be the $xz$-plane, the azimuth angle for $\vec{R}$ vanishes and hence the geometric factor is real. As an example, the values of the geometric factor and its time-variation are shown for the scattering with incident speed $v=0.1$ in Fig. \ref{fig:GeoFactor}. We find that significant transitions occur only when the HI projectile is sufficiently close to the target (note the time for approaching differs with the incident speed), where the field strength is strong and the time-variation of the field is rapid. After the HI passes by, the transition probabilities attenuate asymptotically due to the decreasing geometric factor in the interaction matrix elements.

We note that the asymptotic transition probability of each level does not depend on the incident speed monotonically. This is due to the phase factor in Eq. \eqref{eq:IntMatElem}, which depends on the transition energies. In fact, this phase favors specific transition energies depending on the incident speed. Taking into account the specified transitions included for the current description of the $np$ target (Fig. \ref{fig:SevenBases}), the non-monotonic dependencies of the transition probabilities on the incident speed can be understood. In other words, the transition probability of each state does not necessarily increase with incident speed. For example, the transition probability to the state $^3P_1, M=-1$ is the largest when the incident speed is $v=0.2$. 

In addition, we find that first-order effects increasingly dominate the final state populations as the incident speed of the HI projectile increases. This could be due to the limitation of the current 7-basis system, where higher-lying scattering states are yet to be included. One expects that higher-lying states receive more population as the incident speed increases. Since our main purpose is to define the approach and demonstrate the method of solution, we defer inclusion of a more complete basis to a future effort. 

\subsection{Observables}
\begin{figure}[H]
\centering
\includegraphics[width=16cm]{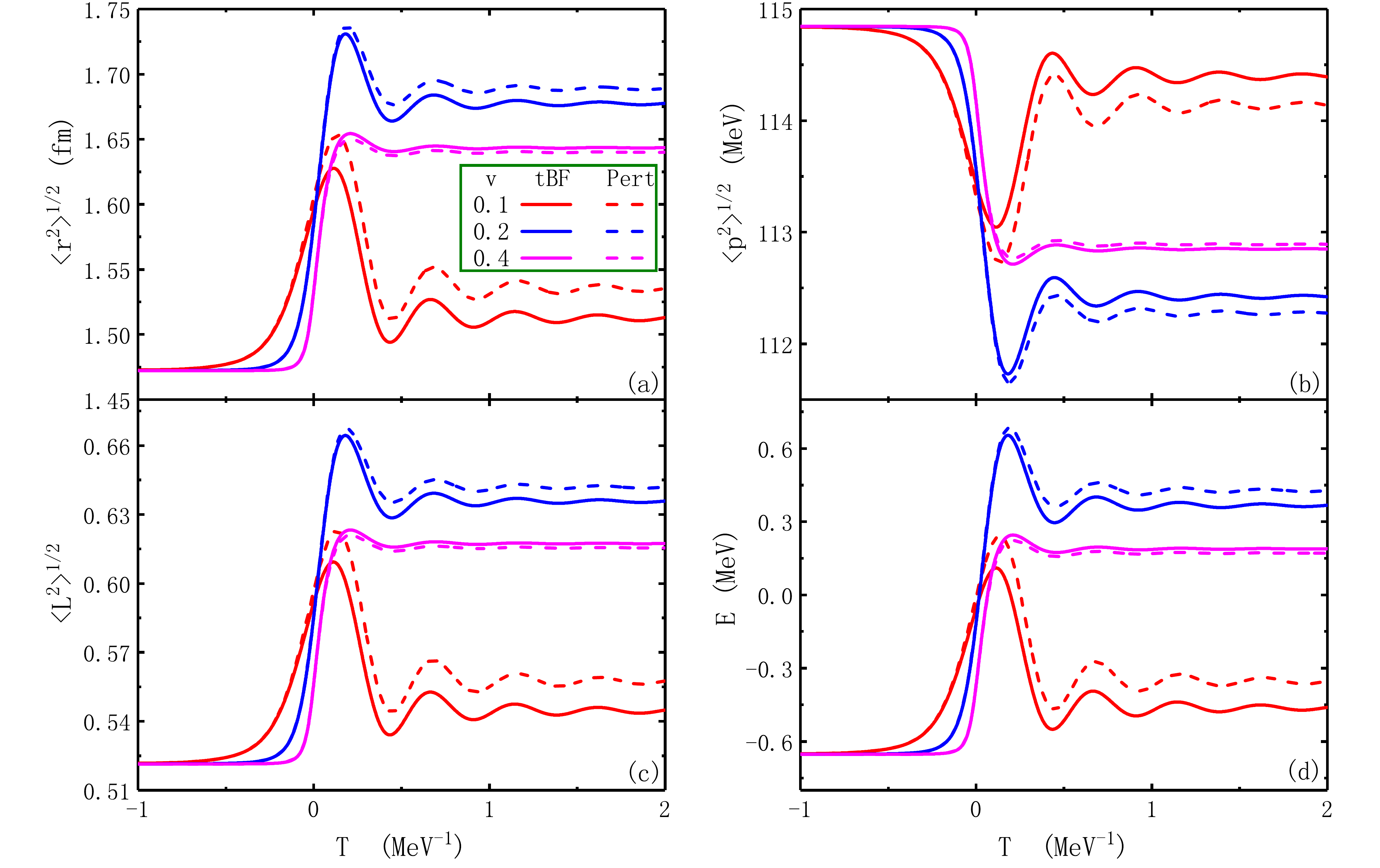}
\caption{Selected observables of the $np$ target as functions of the exposure time and the incident speed of the HI in the middle of the scattering. Panels (a), (b), (c) and (d) show the evolutions of r.m.s. charge radius, r.m.s. momentum, r.m.s. orbital angular momentum and intrinsic energy, respectively.} 
\label{fig:observables}
\end{figure}

With the same simulation conditions as those in Fig. \ref{fig:TransitionProb}, we compute the wave functions of the target during the scattering and evaluate a selected set of operators (we refer to them as ``observables" for brevity) as functions of the exposure time and the incident speed (Fig. \ref{fig:observables}). We again provide calculations based on first-order perturbation theory to compare with those from the non-perturbative tBF method. Note that the expectation values of the observables do not change appreciably until the HI gets sufficiently close to the target, while they relax to respective asymptotic values after the HI flies away from the target. We also comment that the initial values of the observables differ from those for a natural deuteron due to the external HO trap introduced in Eq. \eqref{eq:H0}. For example, the r.m.s. charge radius of the target before the scattering is 1.472 fm, which is about 25$\%$ smaller the experimental measurement 1.975(3) fm for a natural deuteron \cite{Martorell:1995zz,Huber:1998zz}.

All the expectation values of the target observables are evaluated with the time-dependent wave function of the target during the scattering, in which the full quantal coherence is retained. With our limited basis set (Fig. \ref{fig:SevenBases}), the matrix representation of each of our selected operators is diagonal. In other words, the expectation of each observable at a certain moment simplifies here to the calculation of the weighted average (the possible values of the observable weighted by respective eigenbasis probabilities). Therefore, it is not surprising that the evolutions of different observables behave similarly; the time-dependencies of the observables can be easily understood by the results in Fig. \ref{fig:TransitionProb} and by the fact that higher-lying basis states contribute larger r.m.s. charge radii, eigenenergies and r.m.s. orbital angular momenta together with smaller r.m.s. intrinsic momenta. That is, for each observable as a function of the different incident speeds, the sequence of the onsets of the quantum fluctuations in the middle, the subsidence of the oscillations at the end of the scattering, the importance of the higher-order effects and the dependence on the incident speed are easily interpreted in terms of the behaviors of the transition probabilities (Fig. \ref{fig:TransitionProb}). In future applications, with a larger eigenbasis, we anticipate this simple picture will be distorted, for example, by additional coherent effects on the transition matrix elements since the time-dependent amplitude will acquire contributions that are off-diagonal in the eigenbasis.

We find that momentum, angular momentum and energy are transferred significantly to the target when the projectile is near its closest approach. The spikes indicating quantum fluctuations with short-time duration subside as the Coulomb field weakens following the HI's closest approach. After the scattering, we find from Fig. \ref{fig:observables} that the intrinsic motion of the target is excited and that excitation is greater when the incident speed leads to favorable phase coherence within the current level structure. For example, the average intrinsic energies of the scattered $np$ target (panel (d)) increase by at least 0.7 MeV when the incident speeds are 0.2 and 0.4, indicating the important roles of the excited channels. Even for the case with the incident speed $v=0.1$, the average intrinsic energy of the scattered target increases by about 10$\%$.

\begin{figure}[H]
\centering
\includegraphics[width=10cm]{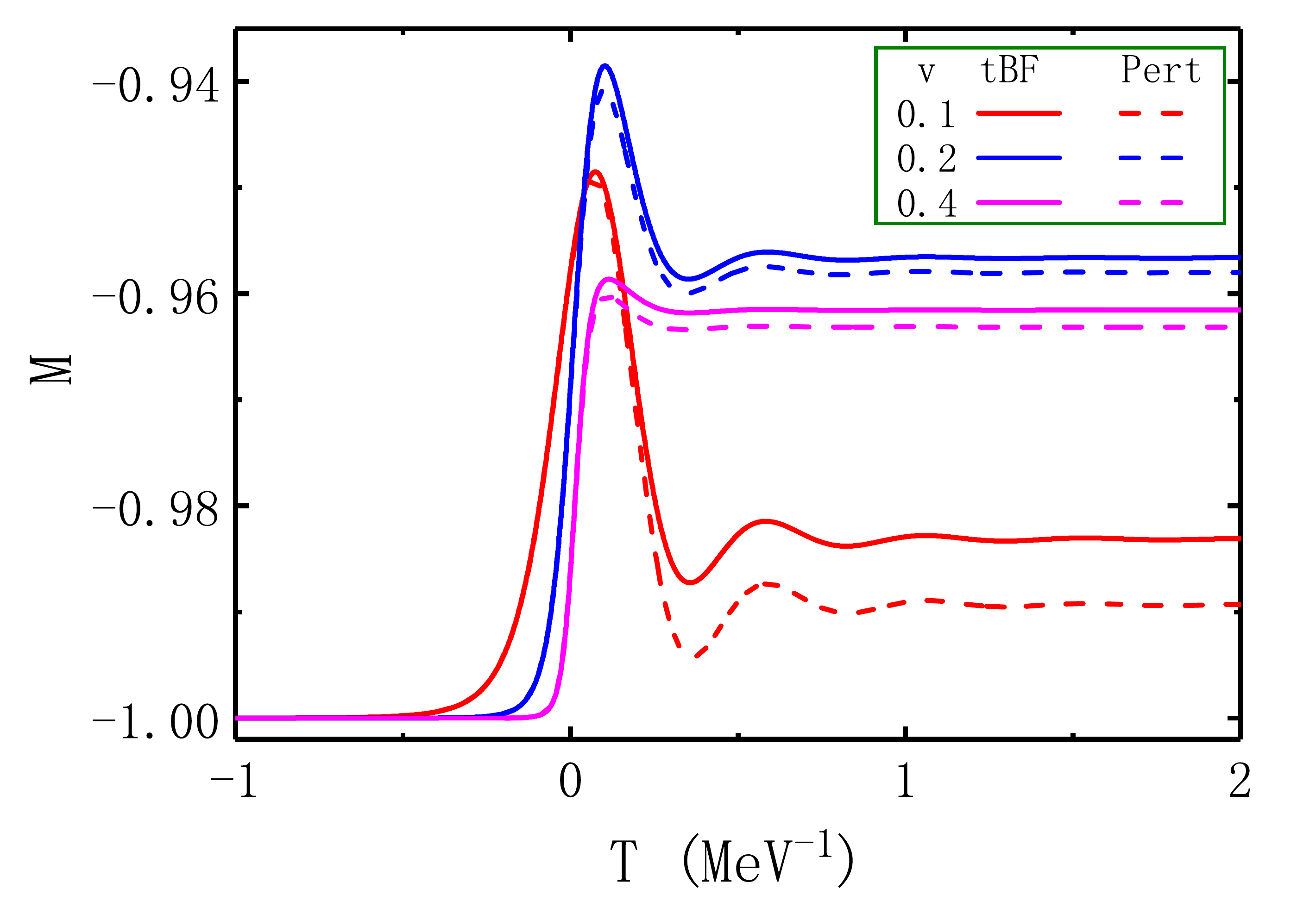}
\caption{Expectation values of the $\hat{z}$-projection of the total angular momentum as functions of the exposure time and the incident speed in the middle of the scattering. These values are calculated in the same manner as the observables in Fig. \ref{fig:observables}. }
\label{fig:MJ}
\end{figure}

We also note that the $\hat{z}$-projection of the total angular momentum, which determines the polarization of the target, is similarly affected during the scattering process as seen in Fig. \ref{fig:MJ}. Indeed, the expectation value of the $\hat{z}$-projection of the total angular momentum indicates the orientation of the target during the scattering.

\subsection{Evolution of the internal charge distribution}

\begin{figure}[H]
\centering
\includegraphics[width=16cm]{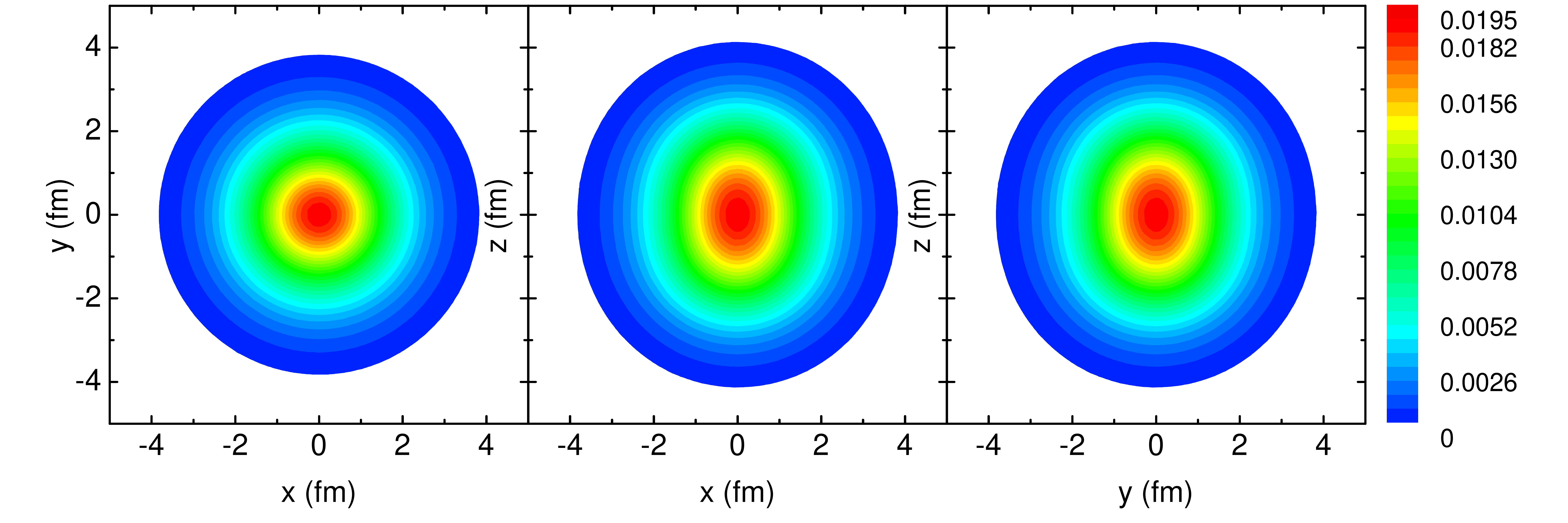}
\caption{The internal charge distribution (in $\text{fm}^{-3}$) of the $np$ target before scattering. The initial target is prepared in the state $(^3S_1,\ ^3D_1),\ M=-1$, in which the polarization is anti-parallel to the $\hat{z}$-axis. The $xz$-plane is the scattering plane (see Fig. \ref{fig:MoonAnd7Pennies}), the $xy$- and $yz$-planes are respectively perpendicular and parallel to the impinging HI.}
\label{fig:InitialDistribution}
\end{figure}

The tBF method enables investigations of the detailed dynamics of the scattering process. As an example, we will show in this work the evolution of the internal charge distribution of the target during the scattering process. Since our main purpose here is to set up the methodology, we shall consider only the case with the incident speed $v=0.1$, where higher-order effects are clearly visible in the complex flow of populations among the levels as discussed above.

\begin{figure}[H]
\centering
\includegraphics[width=17cm]{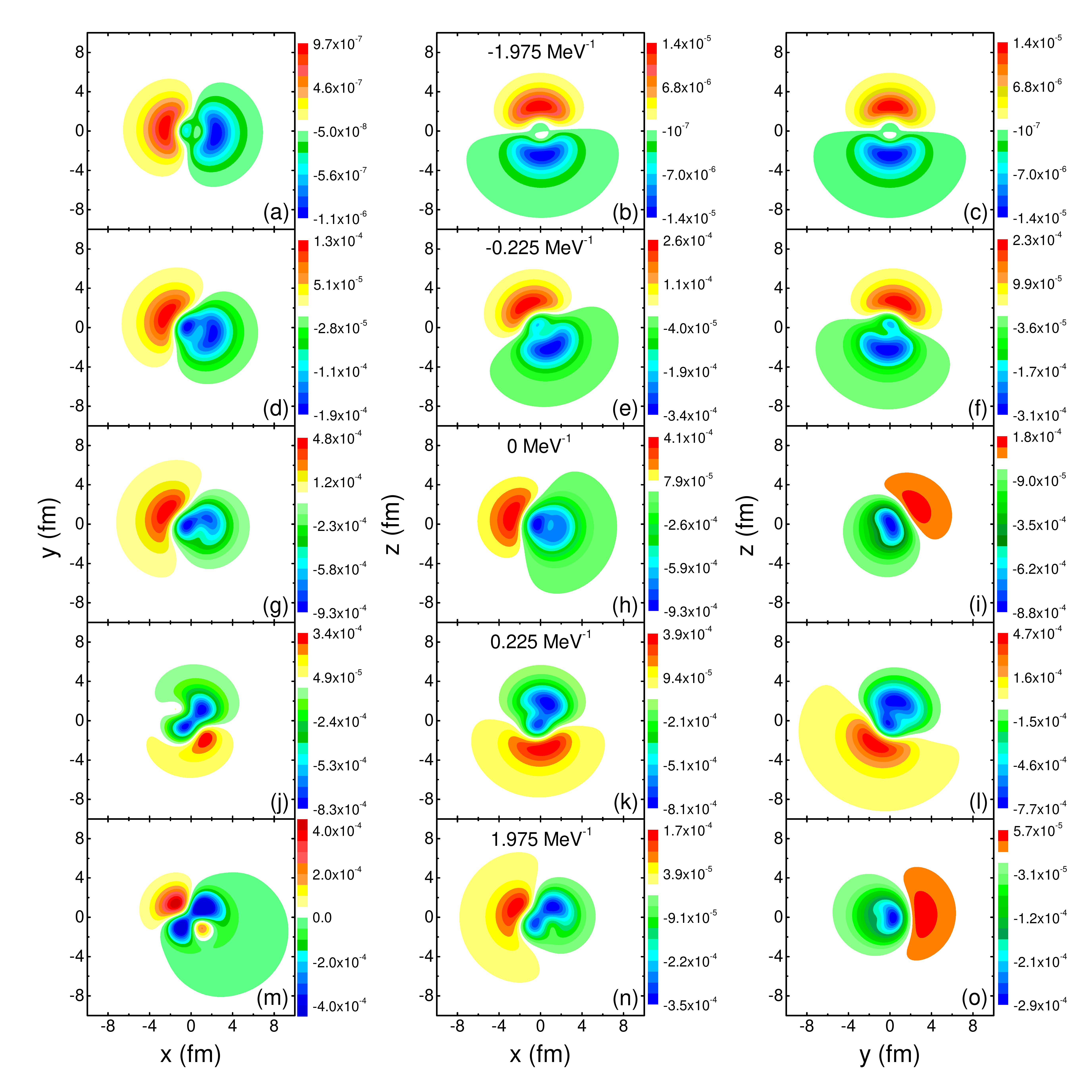}
\caption{The overview of the evolution of the internal charge distribution (in $\text{fm}^{-3}$) of the $np$ target during the scattering. The simulation conditions are the same as those in Fig. \ref{fig:TransitionProb}, except that the incident speed is chosen as 0.1.  These graphs show the difference in the internal charge distributions between the scattered targets at $T=-1.975, -0.255, 0, 0.255, 1.975$ MeV$^{-1}$ and the initial target ($T=-5$ MeV$^{-1}$) in three orthogonal coordinate planes, where the transition amplitudes of each basis state of the target at the selected moments are calculated via the non-perturbative tBF method. See the text for the details.} 
\label{fig:OverViewOfScattering}
\end{figure}

\begin{figure}[H]
\centering
\includegraphics[width=16cm]{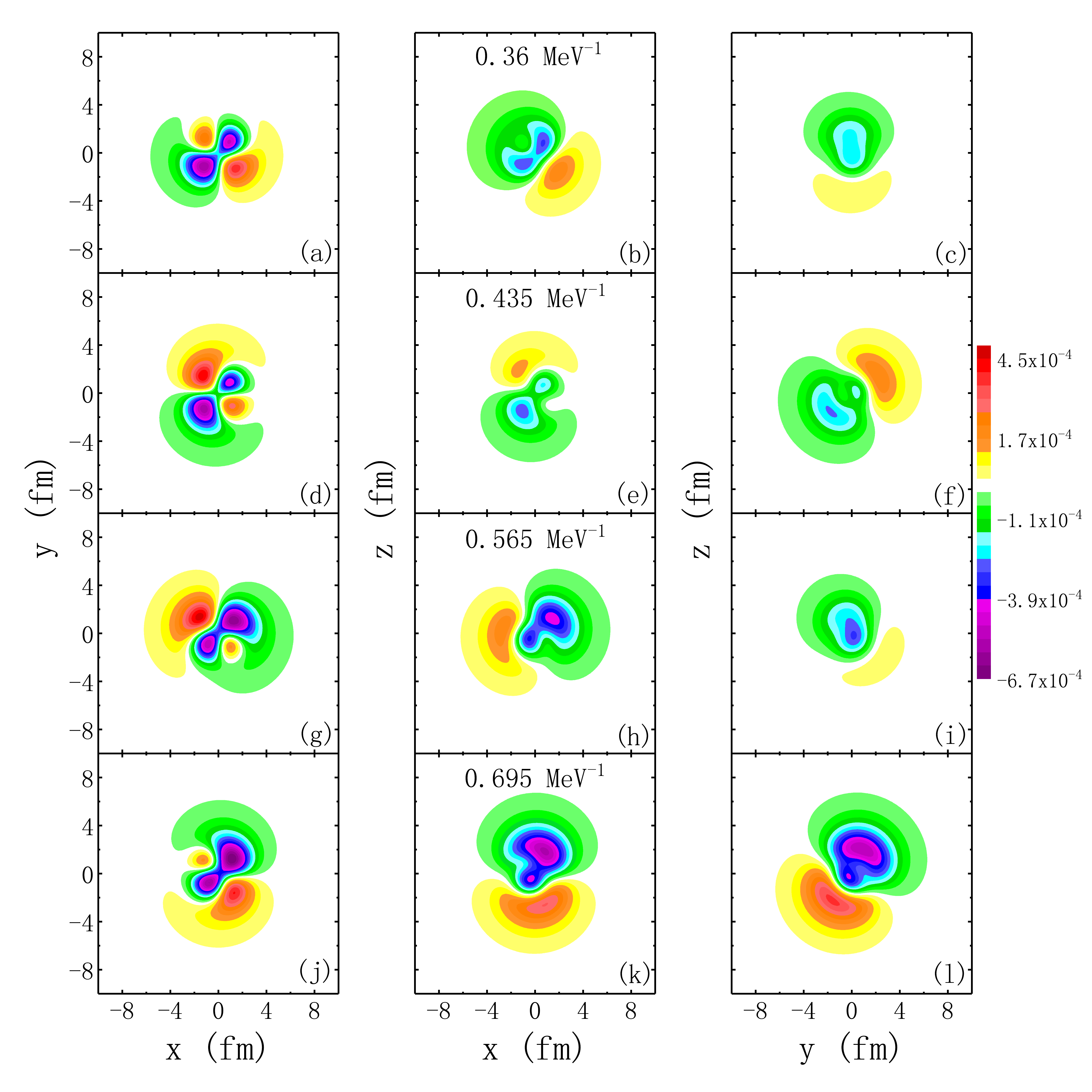}
\caption{Stabilization of the target after the scattering. These graphs show the difference in the internal charge distributions (in $\text{fm}^{-3}$) between  the initial target ($T=-5$ MeV$^{-1}$) and the scattered targets at $T=0.36$ MeV$^{-1}$ (top row), $0.435$ MeV$^{-1}$ (2$^{\text{nd}}$ row), $0.565$ MeV$^{-1}$ (3$^{\text{rd}}$ row), $0.695$ MeV$^{-1}$ (bottom row). The other simulation conditions are the same with Fig. \ref{fig:OverViewOfScattering} and the transition amplitudes of each basis state are calculated via the tBF method. See the text for the details. }
\label{fig:AfterScattering}
\end{figure}

In Fig. \ref{fig:InitialDistribution}, we present the initial internal charge distribution of the target. For the $np$ system under investigation, it is a prolate spheroid with the major axis along the $\hat{z}$-axis. Our distribution differs from the two peaked structure shown in Refs. \cite{Forest:1996kp, MGarcon} due to the fact that our wave functions of the $np$ target are solved implementing the JISP16 $NN$-potential, which is a realistic ``soft" potential without strong short-range correlations. We present the difference in charge distributions between the initial and the scattered targets (Figs. \ref{fig:OverViewOfScattering}, \ref{fig:AfterScattering}) to investigate the dynamics at selected intermediate exposure times. We emphasize that this is the information available within our time-dependent treatment. We provide this information to help develop one's intuition, though it is difficult to imagine an experiment that interrogates for this information.

We find, in general, the scattering of the target can be mainly divided into three sequential stages as described in the following. 

{\bf Stage I:} At the very beginning of the scattering, the internal charge distribution of the target begins to polarize due to the repulsive Coulomb interaction, producing a dumbbell shape (the $1^{\text{st}}$ row of Fig. \ref{fig:OverViewOfScattering}). Shortly thereafter, more of the positive charge density shifts to the far-side (the side away from the HI) of the target, as would be expected from the effect of the repulsive Coulomb force in a classical picture. The dipole fluctuation of the charge density \cite{JEisenberg87V1} is also observed together with the general migration of the positive charge density. As the HI approaches, the amplitude of the dipole fluctuation increases. These oscillations are the result of mixing in the excited states with the initial state, however small those mixings may be.

{\bf Stage II:} As the HI nears the target, the strength of the Coulomb field sensed by the target is stronger and time-variation of the field intensifies. Transitions become stronger (the $2^{\text{nd}}$ row of Fig. \ref{fig:OverViewOfScattering}); the modes of different internal motions become more apparent, generating more complex patterns for the charge distributions. The dipole fluctuation of the charge density is suppressed. The migrated positive charge density (forced by the Coulomb repulsion) still concentrates at the far-side of the target. 

Right after the HI passes its closest approach to the $np$ system, different coordinate planes show the rotational motion of the $np$ target (indicated by the $3^{\text{rd}}$ and $4^{\text{th}}$ rows of Fig. \ref{fig:OverViewOfScattering}). The directions are counter-clockwise in the $xy$- and $xz$-planes and clockwise in the $yz$-plane. These rotational directions are determined by the preparation of the initial target. For example, if we had prepared $(^3S_1,\ ^3D_1),\ M=1$ as the initial state for the target, the direction of the rotation would switch (e.g., the rotation would become clockwise in the $xy$-plane). We verified this by actual simulation. In addition to the rotational motion, fluctuations with complex modes in the charge density occur, as clearly shown in the $4^{\text{th}}$ row of Fig. \ref{fig:OverViewOfScattering}.

{\bf Stage III:} When the HI moves further away, the Coulomb field weakens and its time-variation decreases, reducing the amount of energy, momentum and angular momentum transferred to the target per unit time. The target begins to stabilize. The snapshots for the stabilization process are shown in Fig. \ref{fig:AfterScattering}. Note that we present the sequence of graphs such that the internal charge distribution rotates evenly in the $xy$-plane, as can be easily seen from the steady increase in the azimuth angle of the ``green cloud" in the leftmost column. 

After stabilization (the $5^{\text{th}}$ row of Fig. \ref{fig:OverViewOfScattering}), the scattered target evolves as a superposition of the basis states according to the `unperturbed' Hamiltonian $H_0$. The time evolution of the internal charge distribution of the target shown in Fig. \ref{fig:AfterScattering} repeats, indicating the final state is reached. We find the range of the internal charge distribution of the scattered target expands compared to the initial distribution shown in Fig. \ref{fig:InitialDistribution}. This is signified by  the expansion of the r.m.s. charge radius as shown in the panel (a) of Fig. \ref{fig:observables}. In addition, the complex patterns in the internal charge distribution indicate the excitation of the orbital angular momentum (panel (c) in Fig. \ref{fig:observables}). Finally, we observe the combination of the rotation and oscillations in the charge density, again indicating the excitation of these degrees of freedom.

\section{Conclusions and outlook}
We develop an {\it ab initio}, non-perturbative approach to treat the non-relativistic nuclear structure and scattering problems in a unified manner. We call this approach the time-dependent Basis Function (tBF) method. Within the tBF formalism, the state vector of the system is calculated at the amplitude level during the scattering, by explicitly evaluating the time-evolution operator. The full quantal coherence is therefore retained and we are able to study the detailed dynamics for complex scattering processes. 

As an initial test problem for illustrating the tBF method, we study the Coulomb excitation of a deuteron in a weak harmonic potential (the setup shown in Fig. \ref{fig:MoonAnd7Pennies}). We scatter a $\text{U}^{92+}$ projectile (with the incident speed $v=0.1,\ 0.2,\ 0.4$ and the impact parameter $b=5$ fm) to generate the time-varying Coulomb field, for which the multipole decomposition is performed and only the $E1$ component is kept for illustration. 

In this simple application of the tBF formalism, the structure of the target is solved using the JISP16 $NN$-interaction. With the 3DHO representation, we construct the target Hamiltonian setting the full space truncation parameter $N_{\text{max}} \le 60$ and basis strength $\omega=5$ MeV. To localize the target and to regulate the continuum states, we also introduce a weak HO trap of strength $5$ MeV for the target Hamiltonian. By diagonalization of the target Hamiltonian, the lowest 7 states in the interaction channels $(^3S_1, \ ^3D_1)$, $^3P_0$ and $^3P_1$ are solved (Fig. \ref{fig:SevenBases}). We select these states as a basis set to construct a basis representation for the time-dependent solution of the target under scattering. Note the center of mass excitation of the target is neglected for simplicity. 

Within the basis representation, the time-dependent state vector of the target becomes the wave function, which is made up by a set of amplitudes of respective basis states. Meanwhile, the equation of motion for the scattering takes the form of matrix multiplications. In this work, we prepare the initial target to be polarized in the state $(^3S_1, \ ^3D_1),\ M=-1$ and solve the wave function during the scattering numerically by the MSD2 scheme. In order to reveal the importance of higher-order effects in the scattering, we also solve the wave function via first-order perturbation theory. 

The time-dependent wave function, obtained via either the MSD2 or first-order perturbation theory, is used to investigate the intrinsic excitations of the target. We study the transition probabilities to different basis states as functions of the exposure time and the incident speed. We find that abrupt transitions occur during the scattering, when the strength of the Coulomb field is strong and its time-variation is rapid. The transitions subside and approach asymptotic values as the Coulomb field subsides. 

We also study the feeding of allowed and forbidden states. It is found that higher-order transitions occur later and build smaller populations for the forbidden states. With increasing incident speed, first-order effects dominate the transitions, with non-monotonic dependencies of the transition probabilities on the incident speed clearly visible in the simulations. This could be due to the restricted basis representation for the target in the current model problem.   

The tBF method enables us to study the evolution of the target observables, such as the r.m.s. charge radius, the r.m.s. intrinsic momentum, the r.m.s. angular momentum, the intrinsic energy and the $\hat{z}$-projection of the total angular momentum. Applying the matrix representations of the corresponding operators as well as the time-dependent wave functions, we study the evolution of the observables according to the exposure time and the incident speed. The evolution of these observables is analyzed based on the transition probabilities to different basis states, from which we obtain the transfer of the energy, the momentum and the angular momentum between the background field and the target during the scattering.  

By the tBF method, we expose the dynamics of the scattering directly from the evolution of the internal charge distribution of the target. To illustrate, we show the difference in internal charge distributions between the scattered and initial targets for the case with the incident speed $v=0.1$. We find that the scattering of the target in the time-varying Coulomb field is divided into three sequential stages, i.e., the polarization stage, the transition stage and the stabilization stage. At the end of the scattering, the excitation in the intrinsic degrees of freedom, such as the rotation and fluctuation of the charge density, is evident.

In the future, we will remove the trap and adopt a more complete description for both the interaction channels and the time-varying EM field to investigate the complex processes such as the reorientation \cite{LOu2015, VBar2010, HSeyf2010}, the dissociation \cite{CABert92, LFCanto97,Austern:1087zz} of the deuteron system and the dependence on the $NN$-interaction. The scattering due to the strong nuclear interaction with the heavy ion will also be included in future work. 

\section*{Acknowledgments}
This work was supported in part by the US Department of
Energy (DOE) under Grant Nos.~DE-FG02-87ER40371, DE-SC0018223 (SciDAC-4/NUCLEI) and DE-SC0015376 (DOE Topical Collaboration in Nuclear Theory for Double-Beta Decay and Fundamental Symmetries). A portion of the computational resources were provided by the National Energy Research Scientific Computing Center (NERSC), which is supported by the US DOE Office of Science. Xingbo Zhao is supported by new faculty startup funding from the Institute of Modern Physics, Chinese Academy of Sciences. Peng Yin and Wei Zuo are supported by the National Natural Science Foundation of China (Grant Nos. 11435014, 11705240) and the 973 Program of China (Grant No. 2013CB834405).

\newpage
\section*{Appendix}
We introduce the 3DHO basis in the coordinate representation as
\begin{eqnarray}
\langle \vec{r} | n l S J M \rangle &=& R_{nl}(r) \sum_{m_lm_s} \ (lm_{l}Sm_{s}|J M) Y_{lm_l}(\Omega _{\hat{r}}) \chi_{S m_s} \ \label{eq:CoordinationSpaceWF2}.
\end{eqnarray}
The summations run over all the possible values of the magnetic quantum numbers of the orbital angular momentum $l$ and the spin $S$. $(lm_{l}Sm_{s}|J M)$ denotes the CG-coefficient following the Condon-Shortley convention \cite{JSuhonen}. $ \chi_{S m_s}$ denotes the spin part of the wave function. The radial part of the wave function in the coordinate space \cite{JSuhonen} is
\begin{eqnarray}
R_{nl}(r) &=& \sqrt{\frac{2n!}{r^3_0 \Gamma(n+l+\frac{3}{2})}} \ \Big( \frac{r}{r_0} \Big)^{l} \ \exp\Big[ -\frac{r^2}{2 r_0^2} \Big] \ L_n^{l+\frac{1}{2}}\Big( \frac{r^2}{ r_0^2} \Big) \ \label{eq:RadialWaveFunction} ,
\end{eqnarray}
where $ \Gamma(n+l+3/2)$ is the Gamma function and $L_{n}^{\alpha}({r^2}/{ r_0^2})$ is the associated Laguerre polynomial. For the 3DHO basis in the coordinate representation, the oscillator length is $r_0=\sqrt{1/{m \omega}}$ with $m$ being the reduced mass of the $np$ system and $\omega$ being the oscillator strength. Note that the so-defined $R_{nl}(r)$ starts as positive at the origin. The energy eigenfunction in the coordinate space, which is useful in evaluating Eq. \eqref{eq:densityR}, thus becomes
\begin{eqnarray}
\langle \vec{r} | \beta \rangle &=& \sum_{{n l}} a _{n l} R_{nl}(r) \sum_{m_lm_s} \ (lm_{l}Sm_{s}|J M) Y_{lm_l}(\Omega _{\hat{r}}) \chi_{S m_s} \ \label{eq:CoordinationSpaceWF1} . 
\end{eqnarray}

The matrix element of the operator $\hat{O}$ in the 3DHO representation, $\langle n_jl_jS_jJ_jM_j | \hat{O} | n_kl_kS_kJ_k M_k \rangle$ in Eq. \eqref{eq:3DHOOperator}, is computed in the coordinate space. The results for the operators that are relevant to this work are shown in the following.

The $E1$ matrix element in the 3DHO representation is
\begin{align}
& \langle n_jl_jS_j J_j M_j | \frac{r}{2}Y_{1 \mu} (\hat{r})| n_kl_k S_k J_k M_k \rangle  \nonumber \\
=& \int R_{n_jl_j}^{\ast}(r) \frac{r}{2} R_{n_kl_k}(r) r^2 dr \nonumber \\
& \times \sum_{m_{l_j}m_{s_j}} \sum_{m_{l_k}m_{s_k}}  \delta_{S_jS_k} \delta_{m_{s_j} m_{s_k}} (l_jm_{l_j}S_jm_{s_j}|J_j M_j)(l_km_{l_k}S_km_{s_k}|J_k M_k)  \nonumber \\
& \times (-1)^{m_{l_j}} \sqrt{\frac{3(2l_j+1)(2l_k+1)}{4\pi}}
\begin{pmatrix}
l_j & 1 & l_k \\ 
-m_{l_j} & \mu & m_{l_k}
\end{pmatrix} 
\begin{pmatrix}
l_j & 1 & l_k \\ 
0 & 0 & 0
\end{pmatrix} \ \label{eq:kernel} \ .
\end{align}
In our calculation, we adopt the 3$j$-symbols, e.g.,$$\begin{pmatrix}
l_j & 1 & l_k \\ 
-m_{l_j} & \mu & m_{l_k}
\end{pmatrix} $$ 
following the Condon-Shortley convention \cite{JSuhonen}. The radial integral in Eq. \eqref{eq:kernel} can be computed as 
\begin{eqnarray}
\int R_{n_jl_j}^{\ast}(r) \frac{r}{2} R_{n_kl_k}(r)r^2 dr 
&=& \frac{r_0}{2} \left\{
\begin{array}{c}
\sqrt{n_j+l_j+\frac{3}{2}}\ \delta_{n_jn_k} - \sqrt{n_j} \delta_{n_j,n_k+1} \ \ \ \ \  \text{for} \ l_k=l_j+1  \\ 
\sqrt{n_k+l_k+\frac{3}{2}}\ \delta_{n_jn_k} - \sqrt{n_k} \delta_{n_k,n_j+1} \ \ \ \ \text{for} \ l_j=l_k+1  \\ 
0 \ \ \ \ \ \ \ \ \  \ \ \ \ \ \ \ \ \ \ \ \ \ \ \ \ \ \ \ \ \ \ \ \ \ \ \ \text{else}\  \end{array} \right. \ .
\end{eqnarray}

The matrix element of $r^2$ is
\begin{align}
& <n_jl_jS_jJ_jM_j | r^2 | n_kl_kS_kJ_kM_k > \nonumber \\
=& r_0^2 \  \delta_{l_jl_k} \delta_{S_jS_k} \delta_{J_jJ_k}   \delta_{M_jM_k} \times
\left\{
\begin{array}{c}
2n_j+l_j+\frac{3}{2} \ \  \ \ \ \ \ \ \ \ \ \ \ \ \ \ \ \ \ \text{for} \ n_j=n_k \\
-\sqrt{(n_j+l_j+\frac{3}{2})(n_j+1)} \ \ \ \ \ \ \ \ \text{for} \ n_j=n_k-1 \\
-\sqrt{(n_k+l_k+\frac{3}{2})(n_k+1)} \ \ \ \ \ \ \  \text{for} \ n_j=n_k+1 \\
0 \ \ \ \ \ \ \ \ \ \ \ \ \ \ \ \ \ \ \ \ \ \ \ \text{ else} 
\end{array} \right. \ .
\end{align}

The matrix element of $p^2$ is
\begin{align}
& <n_jl_jS_jJ_jM_j | p^2 | n_kl_kS_kJ_kM_k > \nonumber \\
=& p_0^2 \  \delta_{l_jl_k} \delta_{S_jS_k} \delta_{J_jJ_k}   \delta_{M_jM_k} \times
\left\{
\begin{array}{c}
2n_j+l_j+\frac{3}{2} \ \  \ \ \ \ \ \ \ \ \ \  \ \ \ \ \ \ \ \text{for} \ n_j=n_k \\   
\sqrt{(n_j+l_j+\frac{3}{2})(n_j+1)} \ \ \ \ \ \ \ \ \ \ \text{for} \   n_j=n_k-1 \\
\sqrt{(n_k+l_k+\frac{3}{2})(n_k+1)} \ \ \ \ \ \ \ \ \  \text{for} \   n_j=n_k+1 \\
0 \ \ \ \ \ \ \ \ \ \ \ \ \ \ \ \ \ \ \ \ \ \  \ \text{else}
\end{array} \right. \ ,
\end{align}
where $p_0 = \sqrt{m \omega }$ is the oscillator momentum. 

The matrix element of $L^2$ is
\begin{eqnarray}
\langle n_jl_jS_jJ_jM_j | L^2 | n_kl_kS_kJ_kM_k \rangle &=& {l_j(l_j+1)} \ \delta_{n_jn_k} \delta_{l_jl_k} \delta_{S_jS_k} \delta_{J_jJ_k} \delta_{M_jM_k} \ .
\end{eqnarray}

The matrix element for the $\hat{z}$-projection of the total angular momentum $M$ is
\begin{eqnarray}
\langle n_jl_jS_jJ_jM_j | M | n_kl_kS_kJ_kM_k \rangle &=& M_j \ \delta_{n_jn_k} \delta_{l_jl_k} \delta_{S_jS_k} \delta_{J_jJ_k} \delta_{M_jM_k} \ .
\end{eqnarray}

\newpage

\begin{thebibliography}{100}

\bibitem{Feshbach:1958nx} 
  H.~Feshbach,
  Annals Phys.\  {\bf 5}, 357 (1958).

\bibitem{Feshbach:1962ut} 
  H.~Feshbach,
  Annals Phys.\  {\bf 19}, 287 (1962)
  [Annals Phys.\  {\bf 281}, 519 (2000)].

\bibitem{Witala:2000am} 
  H.~Witala, W.~Gloeckle, J.~Golak, H.~Kamada, J.~Kuros-Zolnierczuk, A.~Nogga and R.~Skibinski,
  Phys.\ Rev.\ C {\bf 63}, 024007 (2001).

\bibitem{Lazauskas:2004hq} 
  R.~Lazauskas and J.~Carbonell,
  Phys.\ Rev.\ C {\bf 70}, 044002 (2004).

\bibitem{Lazauskas:2009gv} 
  R.~Lazauskas,
  Phys.\ Rev.\ C {\bf 79}, 054007 (2009).

\bibitem{Deltuva:2006sz} 
  A.~Deltuva and A.~C.~Fonseca,
  Phys.\ Rev.\ C {\bf 75}, 014005 (2007).

\bibitem{Deltuva:2007xv} 
  A.~Deltuva and A.~C.~Fonseca,
  Phys.\ Rev.\ Lett.\  {\bf 98}, 162502 (2007).

\bibitem{Kievsky:2008es} 
  A.~Kievsky, S.~Rosati, M.~Viviani, L.~E.~Marcucci and L.~Girlanda,
  J.\ Phys.\ G {\bf 35}, 063101 (2008).

\bibitem{Marcucci:2009xf} 
  L.~E.~Marcucci, A.~Kievsky, L.~Girlanda, S.~Rosati and M.~Viviani,
  Phys.\ Rev.\ C {\bf 80}, 034003 (2009).

\bibitem{Quaglioni:2008sm} 
  S.~Quaglioni and P.~Navratil,
  Phys.\ Rev.\ Lett.\  {\bf 101}, 092501 (2008).

\bibitem{Quaglioni:2009mn} 
  S.~Quaglioni and P.~Navratil,
  Phys.\ Rev.\ C {\bf 79}, 044606 (2009).

\bibitem{Navratil:2009ut} 
  P.~Navratil, S.~Quaglioni, I.~Stetcu and B.~R.~Barrett,
  J.\ Phys.\ G {\bf 36}, 083101 (2009).
  
\bibitem{Navratil:2011zs} 
  P.~Navratil and S.~Quaglioni,
  Phys.\ Rev.\ Lett.\  {\bf 108}, 042503 (2012).

\bibitem{Baroni:2012su} 
  S.~Baroni, P.~Navratil and S.~Quaglioni,
  Phys.\ Rev.\ Lett.\  {\bf 110}, no. 2, 022505 (2013).

\bibitem{Baroni:2013fe} 
  S.~Baroni, P.~Navratil and S.~Quaglioni,
  Phys.\ Rev.\ C {\bf 87}, no. 3, 034326 (2013).

\bibitem{Navratil:2016ycn} 
  P.~Navrátil, S.~Quaglioni, G.~Hupin, C.~Romero-Redondo and A.~Calci,
  Phys.\ Scripta {\bf 91}, no. 5, 053002 (2016).

\bibitem{Hagen:2010zz} 
  G.~Hagen, T.~Papenbrock and M.~Hjorth-Jensen,
  Phys.\ Rev.\ Lett.\  {\bf 104}, 182501 (2010).

\bibitem{Hagen:2012sh} 
  G.~Hagen, M.~Hjorth-Jensen, G.~R.~Jansen, R.~Machleidt and T.~Papenbrock,
  Phys.\ Rev.\ Lett.\  {\bf 108}, 242501 (2012).

\bibitem{Hagen:2012rq} 
  G.~Hagen and N.~Michel,
  Phys.\ Rev.\ C {\bf 86}, 021602 (2012).

\bibitem{Papadimitriou:2011jx} 
  G.~Papadimitriou, A.~T.~Kruppa, N.~Michel, W.~Nazarewicz, M.~Ploszajczak and J.~Rotureau,
  Phys.\ Rev.\ C {\bf 84}, 051304 (2011).

\bibitem{Papadimitriou:2013ix} 
  G.~Papadimitriou, J.~Rotureau, N.~Michel, M.~Ploszajczak and B.~R.~Barrett,
  Phys.\ Rev.\ C {\bf 88}, no. 4, 044318 (2013).

\bibitem{Barrett:2015wza} 
  B.~R.~Barrett, G.~Papadimitriou, N.~Michel and M.~Płoszajczak,
  arXiv:1508.07529.

\bibitem{Bang:2000}
J.M. Bang, A.I. Mazur, A.M. Shirokov, Yu.F. Smirnov, S.A. Zaytsev, Ann. Phys. (N.Y.) {\bf 280}, 299 (2000).

\bibitem{Shirokov:2016thl} 
  A.~M.~Shirokov, A.~I.~Mazur, I.~A.~Mazur and J.~P.~Vary,
  Phys.\ Rev.\ C {\bf 94}, no. 6, 064320 (2016).

\bibitem{Shirokov:2016ywq} 
  A.~M.~Shirokov, G.~Papadimitriou, A.~I.~Mazur, I.~A.~Mazur, R.~Roth and J.~P.~Vary,
  Phys.\ Rev.\ Lett.\  {\bf 117}, 182502 (2016).
  
\bibitem{Kravvaris:2017nyj} 
  K.~Kravvaris and A.~Volya,
  Phys.\ Rev.\ Lett.\  {\bf 119}, no. 6, 062501 (2017).

\bibitem{Lynn:2015jua} 
  J.~E.~Lynn, I.~Tews, J.~Carlson, S.~Gandolfi, A.~Gezerlis, K.~E.~Schmidt and A.~Schwenk,
  Phys.\ Rev.\ Lett.\  {\bf 116}, no. 6, 062501 (2016).

\bibitem{Nollett:2006su} 
  K.~M.~Nollett, S.~C.~Pieper, R.~B.~Wiringa, J.~Carlson and G.~M.~Hale,
  Phys.\ Rev.\ Lett.\  {\bf 99}, 022502 (2007).

\bibitem{Rupak:2013aue} 
  G.~Rupak and D.~Lee,
  Phys.\ Rev.\ Lett.\  {\bf 111}, no. 3, 032502 (2013).

\bibitem{Elhatisari:2015iga} 
  S.~Elhatisari, D.~Lee, G.~Rupak, E.~Epelbaum, H.~Krebs, T.~A.~Lähde, T.~Luu and U.~G.~Meißner,
  Nature {\bf 528}, 111 (2015).

\bibitem{Vary:2009gt} 
  J.~P.~Vary {\it et al.},
  Phys.\ Rev.\ C {\bf 81}, 035205 (2010).

\bibitem{Zhao:2013cma} 
  X.~Zhao, A.~Ilderton, P.~Maris and J.~P.~Vary,
  Phys.\ Rev.\ D {\bf 88}, 065014 (2013).

\bibitem{Zhao:2013vga} 
  X.~Zhao, A.~Ilderton, P.~Maris and J.~P.~Vary,
  Proceedings of International Conference ‘Nuclear Theory in the Supercomputing Era — 2013’ (NTSE-2013), Ames, IA, USA, May 13–17, 2013. Eds. A. M. Shirokov and A. I. Mazur. Pacific National University, 
Khabarovsk, Russia, 2014, p. 204; ISBN 978-5-7389-1384-6;
arXiv:1311.2951.

\bibitem{Chen:2017uuq} 
  G.~Chen, X.~Zhao, Y.~Li, K.~Tuchin and J.~P.~Vary,
  Phys.\ Rev.\ D {\bf 95}, no. 9, 096012 (2017).

\bibitem{Du:2017ckx} 
  W.~Du, P.~Yin, G.~Chen, X.~Zhao and J.~P.~Vary,
  Proceedings of the International Conference  `Nuclear Theory in  the Supercomputing Era-2016' (NTSE-2016), Khabarovsk, Russia, September 19-23, 2016. Eds. A. M. Shirokov and  A. I. Mazur.  Pacific National University, Khabarovsk, Russia, 2018, p. 102; arXiv:1704.05520.
  
\bibitem{Alder:1956im} 
  K.~Alder, A.~Bohr, T.~Huus, B.~Mottelson and A.~Winther,
  Rev.\ Mod.\ Phys.\  {\bf 28}, 432 (1956).
  
\bibitem{Winther:1979zz} 
  A.~Winther and K.~Alder,
  Nucl.\ Phys.\ A {\bf 319}, 518 (1979).


\bibitem{JEisenbergV587} J. M. Eisenberg and W. Greiner, {\em Nuclear Theory: Nuclear Models, 3rd. ed.} North-Holland, 1987, p. 584.

\bibitem{AAskar78} A. Askar and A. S. Cakmak, J. Chem. Phys. {\bf 68}, 2794 (1978).

\bibitem{TIitaka94} T. Iitaka, Phys. Rev. E {\bf 49}, 4684 (1994).

\bibitem{ABohr} A. Bohr and B. Mottelson, {\em Nuclear Structure, vol 1, } World Scientific Publishing Co. Pte. Ltd, 1998, p. 92.

\bibitem{JEisenberg88} J. M. Eisenberg and W. Greiner, {\em Nuclear Theory: Excitation Mechanisms of the Nucleus, 3rd. ed.} North-Holland, 1988, p. 235.

\bibitem{JSuhonen} J. Suhonen, {\em From Nucleons to Nucleus: Concepts of Microscopic Nuclear Theory} Springer-Verlag Berlin Heidelberg, 2007.

\bibitem{AShi04} A. M. Shirokov, A. I. Mazur, S. A. Zaytsev, J. P. Vary
and T. A. Weber, Phys. Rev. C {\bf 70}, 044005 (2004).

\bibitem{AShi05} A. M. Shirokov, J. P. Vary, A. I. Mazur, S. A. Zaytsev
and T. A. Weber, Phys. Lett. B {\bf 621}, 96 (2005).

\bibitem{AShi07} A. M. Shirokov, J. P. Vary, A. I. Mazur and T. A.Weber,
Phys. Lett. B {\bf 644}, 33 (2007).

\bibitem{Martorell:1995zz} 
  J.~Martorell, D.~W.~L.~Sprung and D.~C.~Zheng,
  Phys.\ Rev.\ C {\bf 51}, 1127 (1995).
  
\bibitem{Huber:1998zz} 
  A.~Huber, T.~Udem, B.~Gross, J.~Reichert, M.~Kourogi, K.~Pachucki, M.~Weitz and T.~W.~Hansch,
  Phys.\ Rev.\ Lett.\  {\bf 80}, 468 (1998).

\bibitem{Forest:1996kp} 
  J.~L.~Forest, V.~R.~Pandharipande, S.~C.~Pieper, R.~B.~Wiringa, R.~Schiavilla and A.~Arriaga,
  Phys.\ Rev.\ C {\bf 54}, 646 (1996).
  
\bibitem{MGarcon}  
M. Garçon, J.W. Van Orden, {\it The Deuteron: Structure and Form Factors. In: Negele J.W., Vogt E.W. (eds) Advances in Nuclear Physics. Advances in the Physics of Particles and Nuclei, Vol 26}, Springer, Boston, MA, (2001). 
  
\bibitem{JEisenberg87V1} J. M. Eisenberg and W. Greiner, {\em Nuclear Models: Collective and Single-Particle Phenomena, 3rd. ed.} North-Holland, 1987, p. 44.

\bibitem{LOu2015} L. Ou , Z. Xiao, H. Yi, N. Wang, M. Liu, and J. Tian, 
Phys. Rev. Lett. {\bf 115}, 212501 (2015).

\bibitem{VBar2010} V. Baryshevsky, and A. Rouba, Phys. Lett. B {\bf 683}, 299 (2010).

\bibitem{HSeyf2010} H. Seyfarth, R. Engels, F. Rathmann, H. Ströher, V. Baryshevsky, A. Rouba, C. Düweke, R. Emmerich, A. Imig, K. Grigoryev, M. Mikirtychiants, and A. Vasilyev, Phys. Rev. Lett. {\bf 104}, 222501 (2010).

\bibitem{CABert92} C. A. Bertulani and L. F. Canto, Nucl. Phys. {\bf A539}, 163 (1992).

\bibitem{LFCanto97} L. F. Canto, R. Donangelo, A. Romanelli, M. S. Hussein, and A. F. R. de Toledo Piza, Phys. Rev. C {\bf 55}, R570(R) (1997).


\bibitem{Austern:1087zz} 
  N.~Austern, Y.~Iseri, M.~Kamimura, M.~Kawai, G.~Rawitscher and M.~Yahiro,
  Phys.\ Rept.\  {\bf 154}, 125 (1987).

















  


\end{thebibliography}
\end{document}